\documentclass[epj]{svjour}
\usepackage{graphics}
\usepackage{graphicx}
\usepackage{fancyhdr}

\usepackage{amssymb,amsmath}
\usepackage{CJK}
\usepackage{indentfirst}
\usepackage{amsmath}
\usepackage[square, comma, sort&compress, numbers]{natbib}

\begin{document}
\title{ The thermodynamic relationship between the RN-AdS black holes and the RN black hole in canonical ensemble }

\author{Yu-Bo Ma\inst{1,2,3},  Li-Chun Zhang\inst{2,3}, Jian Liu\inst{4},  Ren Zhao\inst{2,3},  Shuo Cao\inst{1}
\thanks{\emph{e-mail:} caoshuo@bnu.edu.cn}%
}                     

\institute{Department of Astronomy, Beijing Normal University,
Beijing 100875, China; \and Institute of Theoretical Physics, Shanxi
Datong University, Datong 037009, China; \and School of Physic,
Shanxi Datong University, Datong, 037009, China; \and School of
Physic, The University of Western Australia,Crawley, WA 6009,
Australia}
%
%
\abstract{In this paper, by analyzing the thermodynamic properties
of charged AdS black hole and asymptotically flat space-time charged
black hole in the vicinity of the critical point, we establish the
correspondence between the thermodynamic parameters of
asymptotically flat space-time and nonasymptotically flat
space-time, based on the equality of black hole horizon area in the
two different space-time. The relationship between the cavity radius
(which is introduced in the study of asymptotically flat space-time
charged black holes) and the cosmological constant (which is
introduced in the study of nonasymptotically flat space-time) is
determined. The establishment of the correspondence between the
thermodynamics parameters in two different space-time is beneficial
to the mutual promotion of different time-space black hole research,
which is helpful to understand the thermodynamics and
quantumproperties of black hole in space-time. }
%
%

\maketitle
\section{Introduction}\label{sec:Intro}

The AdS black hole solution in four-dimensional space-time is an
accurate black hole solution of the Einstein equation with negative
cosmological constant in asymptotic AdS space time \cite{1}. This
solution has the same thermodynamic characteristics as the black
hole solution in asymptotically flat space-time, i.e., the black
hole entropy is equal to a quarter of the event horizon area, while
the corresponding thermodynamics quantity satisfies the law of
thermodynamics of black hole. It is well known that, if taken as a
thermodynamic system, the asymptotically flat black hole does not
meet the requirements of thermodynamic stability due to its negative
heat capacity. However, compared with the asymptotically flat
space-time black hole, the AdS black hole can be in thermodynamic
equilibrium and stable state, because the heat capacity of the
system is positive when the system parameters take certain values.

Therefore, the thermodynamics of AdS charged black holes, in
particular its phase transition in (n+1)-dimensional anti-de Sitter
space-time was firstly discussed and extensively investigated in
Refs.~\cite{3,4}, which discovered the first order phase transition
in the charged non-rotating RN-AdS black hole space-time. Recently,
increasing attention has been paid to the possibility that the
cosmological constant $\Lambda$ could be an independent
thermodynamic parameter (pressure), and the first law of
thermodynamics of AdS black hole may also be established with $P-V$
terms. For instance, the $P-V$ critical properties of AdS black hole
was firstly studied in Ref~\cite{2}, which found that the phase
transition and critical behavior of RN-AdS black hole are similar to
those of the general van der waals-Maxwell system. More
specifically, the RN-AdS black hole exhibits the same $P-V$
(liquid-gas phase transition) critical phase transition behavior and
critical exponent as van der waals-Maxwell system. The phase
transition and critical behavior of various black holes in the
extended phase space of AdS have also been extensively studied in
the literature
\cite{5,6,7,8,9,10,11,12,13,15,16,17,18,19,20,21,22,23,24}, which
showed very similar phase diagrams in different black hole systems.

The asymptotically flat black holes cannot reach thermodynamic
stability, due to the inevitable so-called Hawking radiation. In
order to obtain a better understanding of the thermodynamic
properties and phase transition of black holes, we must ensure that
the black hole can achieve stability in the sense of thermodynamics.
According to the previous results obtained by York et al.(1986)
\cite{25}, achieving thermodynamic stability for asymptotically flat
black hole system also depends on the effect of environments, i.e.,
one needs to consider the ensemble system. Different from the
general thermodynamic system, the self-gravitational system has
inhomogeneity in space, which makes it necessary to determine the
corresponding thermodynamic quantities and their characteristic
values.

The local thermodynamic stability of self-gravitational systems can
be analyzed by considering the extreme value of the Helmholz free
energy of the system. When it reaches to a minimum value, the
corresponding system is at least locally stable. According to the
methods extensively studied in the literature
\cite{26,27,28,29,30,31,32}, the extreme value of the free energy of
gravitational systems can be derived from the action I, i.e., the
partition function of the system at the zero-order approximation can
be calculated by using the Gibbons-Hawking Euclidean action
\cite{33}
\begin{equation}\label{1.1}
Z\approx e^{-I_E}
\end{equation}
Combining this with the Helmbolz free energy $F$ from the equation
$Z=e^{-\beta F }$, we can obtain
\begin{equation}\label{1.2}
I_E(r,T,Q;r_+)=\beta F=\beta E(r,Q;r_+)-S(r_+)
\end{equation}
where $r$, $T$, $Q$ respectively denote the radius, temperature and
electric charge of the cavity, $\beta=1/T$, and $r_+$ is the radius
of the black hole horizon. $E(r,Q;r_+)$ and $S(r_+)$ are the
internal energy and entropy of the black hole in the cavity.
Therefore, when the $r,T,Q$ is determined, the only variable for the
system is $r_+$. The thermal-equilibrium conditions of the black
hole and environment can be determined by the following equation
\begin{equation}\label{1.3}
\left.\frac{d I_E }{d r_+}\right|_{r_+ = \bar{r}_+ }=0
\end{equation}
The conditions that the free energy reaches to its minimum value is
that the system is at local equilibrium state. In order to reach to
the thermal equilibrium, it should satisfy the following criteria
\begin{equation}\label{1.4}
\left.\frac{d^2 I_E }{d r_+^2}\right|_{r_+ = \bar{r}_+ } > 0.
\end{equation}
Utilizing the method described above, the literature
\cite{27,28,29,30,31,32,33} have studied the charged black hole and
black branes, obtained the requirements to meet the thermodynamic
equilibrium conditions, and discussed the phase transition and
critical phenomena. More recently, Eune et al. (2015) investigated
the phase transition based on the corrections of Schwarzschild black
hole radiation temperature \cite{34}. More specifically, both of the
charged black hole and the radiation field outside the black hole
were considered in their work, under the condition that they are
both in the equilibrium state.

On the other hand, Reissner-Nordstrom (RN) black hole and RN-AdS
black hole are the exact solutions of the Einstein equation. The
main difference between AdS space-time and asymptotically flat
space-time is the famous Hawking-Page phase transition \cite{1},
i.e., the AdS background provides a natural ¡°constraint box¡±,
which makes it possible to form a thermal equilibrium between large
stable black hole and hot gas. In the recent study of the phase
transition of the RN black hole, in order to meet the requirement of
thermodynamic stability, one needs to artificially add a cavity
concentric in the horizon of the black hole. However, the
determination of the specific value of the radius of the cavity is
still to be done. In the previous studies of phase transient of RN
and RN-AdS black holes, it was found that both of the two types of
black holes exhibit the same $P-V$ (liquid-gas phase transition)
properties as van der Waals-Maxwell system, which is also consistent
with our finding through the comparison between the phase transition
curves of these two kinds of black holes. Therefore, the consistency
of the thermodynamic stable phase and phase transition between the
cavity asymptotically flat black hole and the black hole in the AdS
space seems to indicate a more profound connotation \cite{26,27}.
One of the roles played by the cavity and the AdS space is to ensure
the conservation of the degree of freedom within a certain system.
Naturally, the discussion of the following problems is the main
motivation of our analysis: Does a duality of a gravitational theory
and a non-gravitational theory exist in the cavity setting? Are
these two types of black holes inherently connected? If so, can the
cavity radius introduced for the RN black hole be determined by the
thermodynamic properties of the RN-AdS black hole?

In this paper, by comparing the phase transition curves of the RN
black hole with the RN-AdS black hole, we establish the equivalent
thermodynamic relations of the two kinds of black holes. We also
discuss the relationship between the radius of cavity introduced
into the RN black hole, the black hole horizon radius, and the
cosmological constant. Finally, we investigate the equivalent
thermodynamic quantities of two kinds of black holes, which provides
theoretical basis for the further exploration of their internal
relation.

\section{Thermodynamic properties of flat space-time charged black hole} \label{sec:1}

To begin with, we review the thermodynamic properties of RN black
holes. The metric of a charged RN black hole is given by
\begin{equation}\label{5}
 d{s^2}=-V(r)d{t^2} + \frac{{d{r^2}}}{{V(r)}} + {r^2}d\Omega _2^2\,
\end{equation}
where
\begin{equation}\label{6}
V(r) = 1 - \frac{{2M}}{r} + \frac{{{{\bar Q}^2}}}{{{r^2}}}.
\end{equation}
The corresponding action expresses as \cite{22,26}
\begin{equation}\label{7}
\begin{aligned}
{I_E}&({r_B},{T_B},\bar Q;{r_ + })={\beta _B}F = {\beta _B}E({r_B},\bar Q;{r_ + }) - S({r_ + })\\
     &={\beta _B}{r_B}\left( {1 - \sqrt {\left( {1 - \frac{{{r_ + }}}{{{r_B}}}} \right)\left( {1 - \frac{{{{\bar Q}^2}}}{{{r_B}{r_ + }}}} \right)} } \right) - \pi r_ + ^2,
\end{aligned}
\end{equation}
Here $T_B$ is the temperature of the cavity and $\beta_B=1/T_B$.
$r_B$ is the radius of concentricity outside a black hole horizon
$r_+$, which can be obtained from the following equation
\begin{equation}\label{8}
V({r_ + }) = 1 - \frac{{2M}}{{{r_ + }}} + \frac{{{{\bar Q}^2}}}{{r_ + ^2}} = 0
\end{equation}
The corresponding reduction quantities are defined as
\begin{equation}\label{9}
{\bar I_E} = \frac{{{I_E}}}{{4\pi r_B^2}},x = \frac{{{r_ + }}}{{{r_B}}},q = \frac{{\bar Q}}{{{r_B}}},\bar b = \frac{{{\beta _B}}}{{4\pi {r_B}}}
\end{equation}
Note that the relation of $q<x<1$ always holds (given $r_+>\bar Q$ ,
$r_B>r_+$) and the reduced action takes the form as
\begin{equation}\label{10}
{\bar I_E}(\bar b,q,x) = \bar b\left( {1 - \sqrt {(1 - x)\left( {1 -
\frac{{{q^2}}}{x}} \right)} } \right) - \frac{1}{4}{x^2}.
\end{equation}
Therefore, we can obtain
\begin{equation}\label{11}
\frac{{d{{\bar I}_x}}}{{dx}} = \frac{{1 - \frac{{{q^2}}}{{{x^2}}}}}{{2{{(1 - x)}^{1/2}}{{\left( {1 - \frac{{{q^2}}}{x}} \right)}^{1/2}}}}\left( {\bar b - {b_q}(x)} \right),
\end{equation}
where the function of the reciprocal of temperature is
\begin{equation}\label{12}
{b_q}(x) = \frac{{x{{(1 - x)}^{1/2}}{{\left( {1 - \frac{{{q^2}}}{x}} \right)}^{1/2}}}}{{1 - \frac{{{q^2}}}{{{x^2}}}}}.
\end{equation}
The condition for the RN Black hole and cavity to reach thermal
equilibrium is
\begin{equation}\label{14}
 \frac{{d{{\bar I}_E}}}{{dx}} = 0  \quad  \Rightarrow   \quad  \bar b = {b_q}(\bar
 x).
\end{equation}
Thus the reciprocal of the black hole radiation temperature can be
written as
\begin{equation}\label{13}
 \frac{1}{{{T_2}}} = 4\pi {r_B}{b_q}(x) = 4\pi {r_B}\frac{{x{{(1 - x)}^{1/2}}{{\left( {1 - \frac{{{q^2}}}{x}} \right)}^{1/2}}}}{{1 - \frac{{{q^2}}}{{{x^2}}}}}.
\end{equation}

The critical charge and critical radius of a black hole can be
determined by the following conditions
\begin{equation}\label{15}
{\left. {\frac{{d{b_q}(x)}}{{dx}}} \right|_{x = {x_c}}} = 0 \quad\quad {\left. {\frac{{{d^2}{b_q}(x)}}{{d{x^2}}}} \right|_{x = {x_c}}} = 0
\end{equation}
Combing Eqs.~(12) and (15), one can easily derive the critical
charge and the critical radius of black hole as \cite{26}
\begin{equation}\label{16}
{q_c} = \sqrt5-2  \quad\quad  {x_c} = 5 - 2\sqrt 5
\end{equation}
The characteristic behavior curve of $b_q(x)$ with respect to $x$ is
shown in Fig.~1.

\begin{figure*}
\begin{center}
\resizebox{0.45\textwidth}{!}{%
  \includegraphics{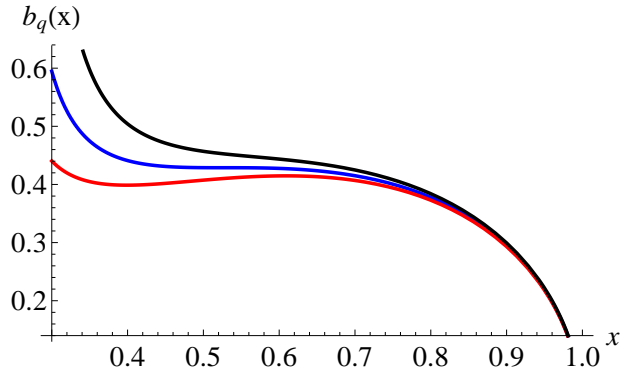} }
\caption{ The characteristic behavior of $b_q(x)$ as a function of
$x$, when $q=q_c -\Delta q (red), q=q_c (blue), q=q_c +\Delta q
(black)$, \textbf{with ${q_c}=\sqrt 5-2$ and $\Delta q=0.03$.} The blue
curve in the middle represent the critical curve.}
\label{fig:1}       
\end{center}
\end{figure*}

\begin{figure*}
\begin{center}
\resizebox{0.45\textwidth}{!}{%
  \includegraphics{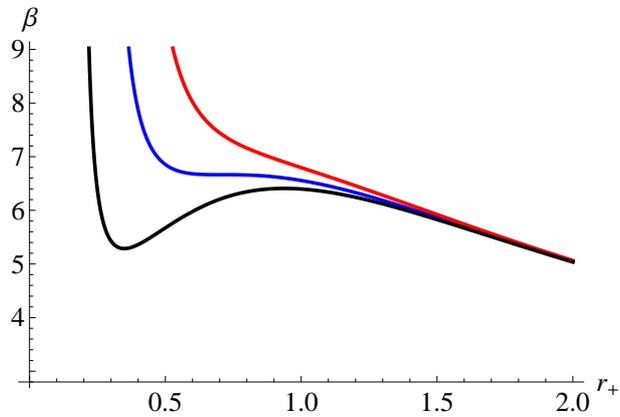}
} \caption{$\beta  - {r_ + }$ diagram when $Q=Q_c$ -\textbf{$\Delta Q (black)$},
$Q=Q_c (blue)$, $Q=Q_c$ +\textbf{$\Delta Q (red)$}, \textbf{with
${Q_c}=\frac{l}{6}$, $l=\sqrt 3$, and $\Delta Q=0.1$.} The blue curve
in the middle represent the critical curve. }
\label{fig:2}       
\end{center}
\end{figure*}

\section{ $P-V$ criticality of charged AdS black holes}\label{sec:2}

To start with, we review some basic thermodynamic properties of the
spherical RN-AdS black holes. In the framework of Schwarzschild-like
coordinates, the metric and the $U(1)$ field read
\begin{equation}\label{17}
d{s^2} =  - f(r)d{t^2} + \frac{{d{r^2}}}{{f(r)}} + {r^2}d\Omega _2^2
\end{equation}
\begin{equation}\label{18}
F = dA, \quad\quad  A =  - \frac{Q}{r}dt
\end{equation}
where $d\Omega _2^2$ stands for the standard element on
\textbf{$S^2$} and the function $f$ is given by
\begin{equation}\label{19}
 f = 1 - \frac{{2M}}{r} + \frac{{{Q^2}}}{{{r^2}}} +
 \frac{{{r^2}}}{{{l^2}}}.
\end{equation}
where $l$ is the AdS length scale and is related to the cosmological
constant as $\Lambda=-3/l^2$. The radius of the black hole horizon,
$r_+$, satisfies the following equation
\begin{equation}\label{20}
  f({r_ + }) = 1 - \frac{{2M}}{{{r_ + }}} + \frac{{{Q^2}}}{{r_ + ^2}} + \frac{{r_ + ^2}}{{{l^2}}} = 0
\end{equation}
and the black hole radiation temperature reads
\begin{equation}\label{21}
{T_1} = \frac{1}{{4\pi {r_ + }}}\left( {1 + \frac{{3r_ +
^2}}{{{l^2}}} - \frac{{{Q^2}}}{{r_ + ^2}}} \right).
\end{equation}
In order to explicitly illustrate the critical phenomenon of RN
black hole, we present the $\beta - {r_ + }$ diagram in Fig.~2, from
which the phase-transition point of system could be obtained as
\begin{equation}\label{22}
 {Q_c} = \frac{l}{6},\quad {r_c} = \frac{l}{{\sqrt 6 }},\quad {\beta _c} = \frac{{\pi l\sqrt 6
 }}{2}.
\end{equation}
These results are well consistent with those obtained in the
previous analysis \cite{2}.

Moreover, the state parameters of a certain system should satisfy
the first law of thermodynamics
\begin{equation}\label{23}
 dM = {T_1}dS + \Phi dQ + VdP
\end{equation}
where the potential $\Phi$, thermodynamic volume $V$ and pressure
$P$ of the black hole respectively express as
\begin{equation}\label{24}
 \Phi  = \frac{Q}{{{r_ + }}}, \quad V = \frac{4}{3}\pi r_ + ^3, \quad P = \frac{3}{{8\pi
 {l^2}}}.
\end{equation}

\section{Thermodynamic relationship of black holes in two different space-time }\label{sec:3}

As can be seen from Fig.~1 and 2, for the two types of black holes,
the curves of the reciprocal of temperature with respect to the
radius of black hole horizon are quite similar to that for the Van
der Waals-Maxwell gas-liquid phase transition (the corresponding
critical exponents can be calculated at the critical point).
Moreover, similar to the cases in the AdS space and dS space, the
specific heat capacity of the black hole in the asymptotically flat
space can be expressed as ${c_v} \sim {(T - {T_c})^{ - 2/3}}$
\cite{30}, where $T_c$ denotes the critical temperature. These
common thermodynamic properties hint us the possibility to make the
thermodynamic properties of the black hole in these space-time well
consistent with each other, only by adjusting some specific
parameters. Therefore, a cavity radius, which may ensure the
thermodynamic equilibrium stability of a system, is introduced in
our discussion of the thermodynamic properties of charged black
holes in the flat space-time. Meanwhile, there is no definite value
for the radius in our analysis, which provides us the possibility to
adjust this parameter to obtain similar thermodynamic properties
from the black holes in two different space-time, and finally derive
the relationship between the radius of the cavity and the radius of
the black hole $r_+$ or the cosmological constant $l$.

It is generally believed that, if the corresponding state parameters
in two thermodynamic system behave the same as the selected
independent variables, these two systems will have the same
thermodynamic properties. In order to obtain similar thermodynamic
properties of the black holes in two different space-time, we ensure
that with the change of horizon radius of black hole, the
temperature and entropy of the two systems are kept to be equal.
That is, the radii of the two black holes are equal to each other in
two different space-time, which leads to the similar entropy in the
two systems. In the vicinity of the critical point, the temperature
of a black hole in the space-time varies synchronously with the
radius of the horizon. Based on the assumption that the entropy and
temperature of a black hole in two different space-time change the
same with the horizon radius, we can obtain the relationship between
the radius of the cavity introduced in the flat space-time and the
radius of the black hole horizon and the cosmological constant.

Supposing the radii of the black hole horizon in the two different
space-time are both $r_+$, which can be respectively determined by
Eq.~(8) and Eq.~(20) for RN black holes and RN-AdS black holes. The
common entropy of these two systems expresses as
\begin{equation}\label{25}
 S = \pi r_ + ^2
\end{equation}
As can be seen from the analysis in Section.~\ref{sec:2}, both of
the two types of black holes show the same characteristics as Van
der Waals-Maxwell system (fluid-gas phase transition), the critical
points of which are respectively given by Eqs.~(16) and (22). It is
apparent that, Eq.~(25) implies the critical point of RN black holes
and RN-AdS black holes are determined when the black hole radius is
the same in two different space-time, while Eq.~(22) shows the
critical value of the RN-AdS black hole is dependent on AdS length
scale $l$. Therefore, when the horizon position $r_+$ and charge $Q$
are given for such black hole, the radiation temperature depends on
the cosmological constant. On the other hand, it can be seen from
Eq.~(13) that, when $q$ and $x$ are fixed, the temperature of RN
black holes depends on the cavity radius $r_B$.

To start with, when the black hole horizon $r_+$ is the same in the
two different space-time, we should determine the requirement of
coincidence of critical curves, which is also the requirement to
make ${T_1} = {T_2}$. With the definition of $y = {r_+ }/l$, the
combination of Eq.~(13) and (21) gives
\begin{equation}\label{26}
 1 - \frac{1}{{36{y^2}}} + 3{y^2} = \frac{{1 - q_c^2/{x^2}}}{{\sqrt {1 - x} \sqrt {1-\frac{{q_c^2}}{x}} }}
\end{equation}
to obtain equal temperature in two time-space, where
${Q_c}=\frac{l}{6}, {q_c}=\sqrt 5-2$. The quantitative relation
between $x$ and $y$ is plotted in Fig.~3. Because the black hole
horizon position $r_+$ is the same for both types of black holes,
the $x-y$ curve also quantifies the relation between $l$ and $r_B$.
We can get $y={y_c}=0.377039$ when $x={x_c}=5-2\sqrt 5$, with which
the critical curves $\beta(x)- x$ of RN black hole and $\beta (y)-x$
of RN-\textbf{AdS} black hole coincide the critical curve of Black Hole.

\begin{figure*}
\begin{center}
\resizebox{0.45\textwidth}{!}{%
  \includegraphics{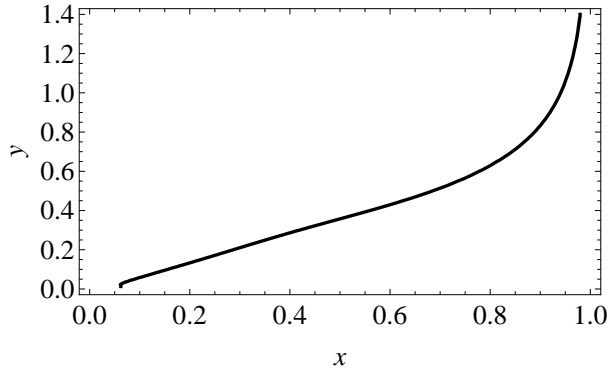}
} \caption{ The $y-x$ curve for Eq.~(26) with ${Q_c}=\frac{l}{6},
{q_c}=\sqrt 5-2, l=\sqrt 3$.}
\label{fig:3}       
\end{center}
\end{figure*}

\begin{figure*}
\begin{center}
\resizebox{0.45\textwidth}{!}{%
  \includegraphics{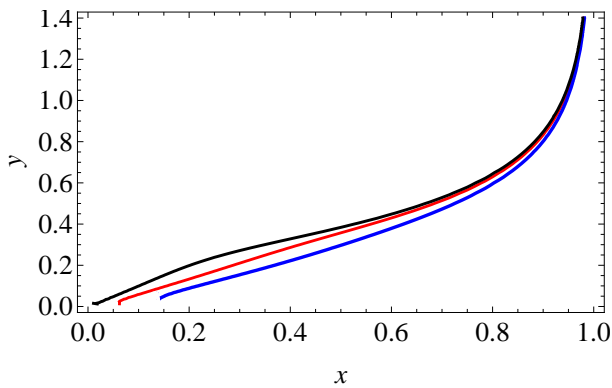}
} \caption{ The $y-x$ curves for $\Delta \tilde Q=0.05$ (blue),
$\Delta \tilde Q=0$ (red), and $\Delta \tilde Q=-0.05 (black)$.}
\label{fig:4}       
\end{center}
\end{figure*}

Secondly, when ${T_1} = {T_2}$, at the critical point $y = {y_c},x =
{x_c}$, supposing there is a deviation of the charge to the
threshold, we can get $\Delta Q = l\Delta \tilde Q$ and $\Delta q$
from the following expression
\begin{equation}\label{27}
 \left({1 + 3\frac{{r_ + ^2}}{{{l^2}}} - \frac{{{Q^2}}}{{r_ + ^2}}} \right) = \frac{{1 - {q^2}/{x^2}}}{{\sqrt {1 - x} \sqrt {1 - \frac{{{q^2}}}{x}} }}
\end{equation}
It is straightforward to obtain
\begin{equation}\label{28}
\begin{aligned}
  - \frac{{2{Q_c}\Delta Q}}{{r_ + ^2}}&=- \frac{{\Delta \tilde Q}}{3{y_c^2}}\\
                                       &=\frac{1}{{\sqrt {1 - {x_c}} {{\left( {1 - \frac{{{q^2}}}{{{x_c}}}} \right)}^{3/2}}}}\frac{{{q_c}}}{{{x_c}}}\left( {1 - \frac{2}{{{x_c}}} + \frac{{q_c^2}}{{x_c^2}}} \right)\Delta q
\end{aligned}
\end{equation}
Substituting ${y_c} = 0.377039$, ${x_c} = 5 - 2\sqrt 5$ into the
above equation, we can get the relationship between $\Delta \tilde
Q$ and $\Delta q$ and the Eq.~(27) can be rewritten as
\begin{equation}\label{29}
 \left( {1 + 3{y^2} - \frac{1}{{36{y^2}}} - \frac{{\Delta \tilde Q}}{{3{y^2}}}} \right) = \frac{{1 - {q^2}/{x^2}}}{{\sqrt {1 - x} \sqrt {1 - \frac{{{q^2}}}{x}} }}
\end{equation}
When $\Delta \tilde Q$ is given, one can get the value of
\textbf{$\Delta q$} from Eq.~(28). Denoting $q={q_c}+\Delta q=\sqrt
5-2+\Delta q$ in Eq.~(29), we can obtain the $y-x$ curve deviating
from the critical charge. The results are plotted in Fig.~4, in
which the blue, red and black lines respectively correspond to the
three cases with the critical charge, above the critical charge and
under critical charge. It is obvious that when the relation between
$y$ and $x$ satisfies Eqs.~(26) and (29), the critical curves $\beta
(x) - x$ of RN black holes and $\beta (y) - y$ of RN-AdS black holes
coincide, which implies the equal temperature of the two different
space-time. Therefore, similar to the definition of entropy in
Eq.~(25), the heat capacity of the two systems can also be written
as
\begin{equation}\label{30}
  {C_Q} = {T_1}{\left( {\frac{{\partial {S_1}}}{{\partial {T_1}}}} \right)_Q} = {C_{\bar Q}} = {T_2}{\left( {\frac{{\partial {S_2}}}{{\partial {T_2}}}} \right)_{\bar Q}}
\end{equation}

We remark here that in the discussion above, the cavity radius is
introduced as a state parameter to test the thermodynamic properties
near the critical point of the RN black hole in the asymptotically
flat space-time. This procedure follows the study of the
thermodynamic properties near the critical point of the RN-AdS black
hole, where the cosmological constant is taken as the state
parameter in the thermodynamic system. Our results demonstrates that
the change of the two systems is the same near the critical point,
as long as the variables $y$ and $x$ introduced in the two system
satisfy Eqs.~(26) and (29). As is shown in Figs.~1-2, the radius of
the cavity is a one-to-one correspondence with the cosmological
constant, and thus could be the dual of the cosmological constant in
non-asymptotic flat space-time.

\section{Conclusion} \label{sec:4}

It is well known that black hole is an ideal system to understand
the nature and behavior of quantum gravity. On one hand, black hole
provides an ideal model to study all kinds of interesting behaviors
of classical gravitation (under the sense of general relativity); on
the other hand, it can be regarded as a macroscopic quantum system
with unique thermodynamic properties (the entropy, temperature and
holographic properties of gravitation are quantum), which provides
an important window to probe the quantum gravity. More importantly,
a better understanding of the black hole singularity, cosmological
singularity, and cosmological inflation needs a basic theory of
space-time gravitation. Up to now there is still no mature theory to
precisely describe the quantum characteristics of the asymptotic
flat black holes and non-asymptotically flat space-time black holes.
However, string theory firstly provided the microcosmic explanations
for the entropy of some AdS black holes, which predicted the
existence of M-theory and some duality relations (especially AdS /
CFT correspondence) and realized the holographic properties of
gravitational systems \cite{36}.

In this paper, by setting the black-hole horizon at the same value
and assuming the temperature near the critical point is equal in two
space-time, we have discussed the relationship between the radius of
the cavity $r_B$ and the cosmological constant $l$. The former
parameter is always introduced in studying the thermodynamics of the
charged black hole in the flat space-time, while the latter is
related to the non-flat space-time. Our results may provide a
theoretical basis for exploring the internal relations of
thermodynamics in the black space. Moreover, by establishing the
correspondence between the thermodynamic parameters of black holes
in the asymptotically flat space-time and non-asymptotically flat
space-time, we hope to seek the correspondence between the quantum
properties of the black hole in two different space-time, which is
beneficial to understand the thermodynamics and quantum properties
of black holes in different space-time.


\section*{Acknowledgments}

The authors declare that there is no conflict of interest regarding
the publication of this paper. This work was supported by the Young
Scientists Fund of the National Natural Science Foundation of China
(Grant Nos.11605107 and 11503001), in part by the National Natural
Science Foundation of China (Grant No.11475108), Supported by
Program for the Innovative Talents of Higher Learning Institutions
of Shanxi, the Natural Science Foundation of Shanxi
Province,China(Grant No.201601D102004) and the Natural Science
Foundation for Young Scientists of Shanxi Province,China (Grant
No.201601D021022), the Natural Science Foundation of Datong
city(Grant No.20150110).

\end{document}